\documentclass[final,onecolumn,10pt]{IEEEtran}
\usepackage{cite}
\usepackage{bm}
\usepackage{url}
\usepackage[usenames,dvipsnames]{color}
\usepackage[dvipsnames]{xcolor}
\usepackage{amsmath}
\usepackage{mathtools}
\usepackage[]{amssymb}
\usepackage{amsthm}
\usepackage{float}
\usepackage{dsfont}
\usepackage{array}
\usepackage{booktabs}
\usepackage{multirow}
\usepackage{footmisc}

\usepackage{graphicx}
\usepackage[para,flushleft]{threeparttable} 
\usepackage{multirow}	
\usepackage{dcolumn} 

\usepackage[ruled]{algorithm2e}
\usepackage{xifthen} 
\usepackage{enumitem}

\makeatletter
\newcommand{\removelatexerror}{\let\@latex@error\@gobble}
\makeatother

\newif \ifwebcolor
\webcolortrue

\def \ba {\begin{array}}
\def \ea {\end{array}}
\def \benu {\begin{enumerate}}
\def \eenu {\end{enumerate}}
\def \bdes {\begin{description}}
\def \edes {\end{description}}

\def \bitem {\begin{itemize}}
\def \eitem {\end{itemize}}

\def \bfl {\begin{flushleft}}
\def \efl {\end{flushleft}}
\def \bfr {\begin{flushright}}
\def \efr {\end{flushright}}

\def \beq {\begin{equation}}
\def \eeq {\end{equation}}
\def \bqa {\begin{eqnarray}}
\def \eqa {\end{eqnarray}}
\def \bqa* {\begin{eqnarray*}}
\def \eqa* {\end{eqnarray*}}

\def \bal {\begin{align}}
\def \eal {\end{align}}

\begin{document}
%
\title{
High Accuracy Distributed Kalman Filtering for Frequency and Phase Synchronization\\in Distributed Phased Arrays}

\author{\IEEEauthorblockN{Mohammed Rashid, ~\IEEEmembership{Member,~IEEE}, and 
Jeffrey A. Nanzer, ~\IEEEmembership{Senior Member,~IEEE}}
\thanks{Manuscript received 2023.}
\thanks{This work has been submitted to the IEEE for possible publication. Copyright may be transferred without notice, after which this version may no longer be accessible.}
\thanks{This work was supported in part by the Office of Naval Research under grant number N00014-20-1-2389. Any opinions, findings, and conclusions or recommendations expressed in this material are those of the author(s) and do not necessarily reflect the views of the Office of Naval Research. \textit{(Corresponding author: Jeffrey A. Nanzer.)}}
\thanks{The authors are with the Electrical and Computer Engineering Department, Michigan State University, East Lansing, MI 48824 (e-mail: \mbox{rashidm4@msu.edu,} nanzer@msu.edu).}}

\maketitle
\thispagestyle{empty}
\pagestyle{empty}

\def\bda{\mathbf{a}}
\def\bdd{\mathbf{d}}
\def\bde{\mathbf{e}}
\def\bdf{\mathbf{f}}
\def\bdg{\mathbf{g}} 
\def\bdh{\mathbf{h}}
\def\bdm{\mathbf{m}}
\def\bds{\mathbf{s}} 
\def\bdn{\mathbf{n}}
\def\bdu{\mathbf{u}}
\def\bdv{\mathbf{v}}
\def\bdw{\mathbf{w}} 
\def\bdx{\mathbf{x}} 
\def\bdy{\mathbf{y}} 
\def\bdz{\mathbf{z}}
\def\bdA{\mathbf{A}}
\def\bdC{\mathbf{C}}
\def\bdD{\mathbf{D}} 
\def\bdF{\mathbf{F}}
\def\bdG{\mathbf{G}} 
\def\bdH{\mathbf{H}}
\def\bdI{\mathbf{I}}
\def\bdJ{\mathbf{J}}
\def\bdU{\mathbf{U}}
\def\bdX{\mathbf{X}}
\def\bdK{\mathbf{K}}
\def\bdQ{\mathbf{Q}}
\def\bdR{\mathbf{R}}
\def\bdS{\mathbf{S}}
\def\bdV{\mathbf{V}}
\def\bdW{\mathbf{W}}
\def\bdGamma{\bm{\Gamma}}
\def\bdgamma{\bm{\gamma}}
\def\bdalpha{\bm{\alpha}}
\def\bdmu{\bm{\mu}}
\def\bdSigma{\bm{\Sigma}^n}
\def\bdOmega{\bm{\Omega}}
\def\bdxi{\bm{\xi}}
\def\bdl{\bm{\ell}}
\def\bdLambda{\bm{\Lambda}}
\def\bdeta{\bm{\eta}}
\def\bdPhi{\bm{\Phi}}
\def\bdpi{\bm{\pi}}
\def\bdtheta{\bm{\theta}}
\def\bdTheta{\bm{\Theta}}
\def\bddelta{\bm{\delta}}

\def\btau{\bm{\tau}}
\def\deg{\circ}

\def\tq{\tilde{q}}
\def\tbdJ{\tilde \bdJ}
\def\l{\ell}
\def\bdzero{\mathbf{0}} 
\def\bdone{\mathbf{1}} 
\def\Exp{\mathbb{E}} 
\def\exp{\text{exp}} 
\def\ra{\rightarrow}

\def\R{\mathbb{R}} 
\def\C{\mathbb{C}} 
\def\CN{\mathcal{CN}} 
\def\N{\mathcal{N}} 
\begin{abstract}
Precise frequency and phase synchronization are among the important aspects in a coherent distributed phased array antenna system, and are among the most challenging to achieve for microwave frequencies and above. 
We propose a high accuracy distributed Kalman filter (HA-DKF) in which 
the nodes in the array perform consensus averaging on their locally shared measurements, the innovation noise covariances, and the KF-predicted as well as the KF-updated estimates and error covariances, yielding improved electrical state synchronization compared to other works in the literature. We evaluate the algorithm through simulation and show that it converges in fewer iterations and yields better electrical state synchronization than other algorithms.

\end{abstract}


\section{Introduction}\label{Intro-section}


A distributed phased array is
a coherent antenna array system implemented 
by coordinating multiple spatially distributed antenna systems (nodes) 
at the level of the radio frequency (RF) wavelength, mimicking the functionality of a large single-platform phased array. Distributed arrays can be used in several applications including distributed radar~\cite{dautoradar_1, Chen_2022} and distributed massive MIMO systems~\cite{Distd_mMIMO,Deng_2022}, and may be implemented in closed-loop~\cite{DCB_2013,1_bit_feedback,3_bit_feedback,retrodirective_2016} or 
open-loop architectures~\cite{Nanzer_Survey_2021}. 
In a closed-loop distributed array, the electrical states of the nodes are synchronized based on the feedback from the beamforming destination, whereas 
in an open-loop array, the nodes only rely on the inter-node coordination to 
achieve the synchronized state of the array. Thus, open-loop distributed array 
can be used for both wireless communication and radar applications, as the needed 
feedback is not feasible in the sensing and receive-only operations. 

For open-loop distributed phased array, centralized coordination approaches have been demonstrated previously~\cite{OCDA_2017, Serge_Access_2021}, however such approaches are challenging to scale and suffer from single points of failure. 
Decentralized consensus-based algorithms are more robust to node failures and are more easily scalable~\cite{Hassna_TWC,rashid2022frequency}. Consensus-based algorithms rely only on iteratively broadcasting the electrical state information between the neighboring nodes, where each node computes an average of its local state and that of its neighbors to update its state in each iteration. This approach ensures that the array is more robust to node failures and enables direct scalability.

Our prior work has investigated the use of a Kalman filtering (KF) based decentralized 
frequency and phase consensus (KF-DFPC) algorithm to predict the temporal variation of the frequency and phase drift in a distributed phased array and thereby reduce the residual phase error of the array~\cite{rashid2022frequency}. In this algorithm, the nodes iteratively share their KF-updated frequency 
and phase estimates as well as the error covariances with their neighbors, and compute a weighted average of the shared values in each iteration to reach a consensus, which is referred to here as a \textit{consensus on estimates and error covariances} (CEEC) approach. 
Other approaches can be characterized as \textit{consensus on estimates} (CE), 
in which the nodes only share their KF-updated estimates of the electrical 
states with their neighbors~\cite{Xin_2022}; \textit{consensus on measurements} (CM), where only local measurements are shared~\cite{Saber_2007, Kamgarpour_2008, Sayed_2010}; or \textit{consensus 
on information} (CI), where the KF-predicted information matrix and information vector are shared~\cite{BATTISTELLI_2014,HCMCI_2015}.

In this paper, we propose a high accuracy distributed Kalman filtering (HA-DKF) algorithm that combines the CM, CI, and CEEC approaches.
Essentially, all CM, CI, and CEEC schemes are complementary, and thus their combination in a single algorithm leads to improved performance in distributed electrical state coordination.
This is because upon synchronization 
the physical states of all the nodes are evolving following a globally varying physical process. Thus, at that stage, there exist a 
correlation between the states of the different 
nodes which is unknown and difficult to compute in a distributed network. 
However, to account for the unknown correlation, 
fusing the KF-predicted and the KF-updated information matrices in the CI and CEEC schemes, respectively, 
introduces unanimity between the nodes in the state estimation that improves the synchronizing performance of HA-DKF. 
Moreover, locally fusing the KF-updated information vectors 
and the KF-updated information matrices in each iteration of HA-DKF under the CEEC scheme provides a better averaged state estimate and thus a better a priori distribution 
for the next iteration of local Kalman filtering that improves the residual phase 
error upon convergence. 
The performance of the proposed HA-DKF algorithm is validated through simulations and it is shown that HA-DKF performs significantly better than the other filtering algorithms in reducing the residual 
phase error and the number of iterations required for the convergence. 

\section{Frequency and Phase Modeling in\\Distributed Phased Arrays}\label{sys_mod}

We consider a distributed phased array with $N$ nodes that are coordinating with 
each other over bi-directional communication links to synchronize their generated signals and thus perform coherent beamforming toward a targeted direction. 
This network of nodes can be represented by a graph 
$\mathcal{G}=(\mathcal{V},\mathcal{E})$ wherein $\mathcal{V}=\{1,2,\ldots,N\}$ denotes the set of vertices (nodes), 
and $\mathcal{E}=\{(n,m)\colon n,m\in \mathcal{V}\}$ is the set of 
undirected edges or communication links between the nodes. 
Let the signal generated by node $n$ in 
the $k$-th iteration is written as 
$s_n(t)=e^{j\left(2\pi f_n(k) t+\theta_n(k)\right)}$ for $t\in[(k-1)T,kT]$ 
and $k\in\{1,2,\ldots\}$. The parameter $T$ represents 
the time duration of the signal, whereas $f_n(k)$ and 
$\theta_n(k)$ denote the frequency and phase of the node in the $k$-th iteration, respectively. 
In a distributed consensus averaging algorithm, the nodes iteratively exchange their frequencies and phases  
with their neighboring nodes, and update these parameters in every iteration 
by computing a weighted average of the shared values 
to reach a consensus. Each node in 
a distributed array has its own local oscillator, 
and these oscillators undergo random frequency drift, phase drift, 
and phase jitter in between the update intervals. Thus, the frequency and phase of node $n$ in the $k$-th iteration can be written as  
\begin{align}\label{FreqPhase_model}
f_n(k)&=f_n(k-1)+\delta f_n\nonumber\\
\theta_n(k)&=\theta_n(k-1)+\delta\theta^f_n+\delta\theta_n,
\end{align}
in which $\delta f_n$ is the frequency drift of the oscillator at node $n$ which 
is normally distributed as $\N\left(0,\sigma^2_f\right)$; the 
standard deviation $\sigma_f$ represents the Allan deviation (ADEV) 
of an oscillator which is modeled 
as $\sigma_f=f_c\sqrt{\frac{\beta_1}{T}+\beta_2 T}$ 
with $\beta_1=\beta_2=5\times 10^{-19}$ representing the design parameters of a quartz crystal oscillator 
\cite{Hassna_TWC, rashid2022frequency}; the phase $\delta\theta^f_n$ 
in \eqref{FreqPhase_model} denotes the phase drift 
due to the temporal variation of the frequency 
drift $\delta f_n$ between the update intervals, and it is computed by \mbox{$\delta\theta^f_n=-\pi T\delta f_n$}~\cite{rashid2022frequency};  
finally, the phase $\delta\theta_n$ in \eqref{FreqPhase_model} represents the phase 
jitter of an oscillator which is modeled as 
$\delta\theta_n\sim \N(0,\sigma_\theta)$. The standard deviation $\sigma_\theta$ 
can be mathematically modeled as 
$\sigma_\theta=\sqrt{2\times 10^{A/10}}$ in which $A$ is the integrated phase noise power of an oscillator 
that can be computed from its phase noise profile. In this work, we set 
$A=-53.46$ dB to model a typical high phase noise voltage controlled oscillator  
\cite{Serge_Access_2021, rashid2022frequency}. To initialize the 
process in \eqref{FreqPhase_model}, 
we assume that 
$f_n(0)\sim\N(f_c,\sigma^2)$, in which $f_c$ is the nominal carrier frequency and 
$\sigma=10^{-4}f_c$ denotes a crystal clock accuracy of $100$ parts per million (ppm), whereas $\theta_n(0)\sim\mathcal{U}(0,2\pi)$ which represents the 
initial phase offset due to the hardware and the oscillator.

Due to the time-varying frequencies and phases of the oscillators, we assume that 
the nodes iteratively estimate these parameters from the observed signals 
to update them in each iteration using a weighted average. 
Thus, the estimated 
frequency and phase of node $n$ in the $k$-th iteration can be given by 
\begin{align}\label{FreqPhase_est}
\hat{f}_n(k)=f_n(k)+\varepsilon_f\nonumber\\
\hat{\theta}_n(k)=\theta_n(k)+\varepsilon_\theta,
\end{align}
where $\varepsilon_f$ and 
$\varepsilon_\theta$ are the frequency and phase estimation errors in the 
$k$-th iteration that are 
normally distributed with zero mean and standard deviation $\sigma^m_f$ and 
$\sigma^m_\theta$, respectively. These standard deviations are set 
equal to their Cramer Rao lower bounds (CRLBs) 
as $\sigma^m_f=f_c\sqrt{\frac{6}{(2\pi)^2 L^3 \text{SNR}}}$ and 
$\sigma^m_\theta= \frac{2L^{-1}}{ \text{SNR}}$ \cite{Richards_radar}. 
In these equations, the parameter $L=Tf_s$ denotes 
the number of samples of the observed signals collected over the time period $T$ 
with sampling frequency $f_s$, 
and the quantity SNR represents the signal-to-noise ratio of the observed signals. 
In the following, 
we describe our proposed high accuracy distributed Kalman filtering algorithm 
to synchronize the frequencies and phases of the nodes in a distributed array 
by only local information sharing, Kalman filtering, and consensus averaging. 

\section{High Accuracy Distributed Kalman Filtering}\label{filtering_sectn}

\subsection{Distributed Kalman Filtering}
In a distributed phased array, the temporal variation in the electrical states of a node can 
be modeled using a state-space model wherein the process noise at the update 
time can be described by the oscillator-induced frequency and phase offset errors, 
and the measurement noise can be defined using the estimation errors of the electrical states. 
Using this approach, a distributed Kalman filtering based KF-DFPC algorithm was proposed in \cite{rashid2022frequency}, where it was shown that the use of local KF with consensus averaging 
significantly reduces 
the residual phase error of the array, 
particularly when fast update rates are used for 
the array synchronization\footnote{In distributed phased arrays, typically 
the update rate is on the order of Hz to kHz, which is required to 
avoid larger oscillator drifts and mitigate any decoherence due to the platform 
vibration and motion~\cite{Serge_Access_2021, Pratik_2017}.}. 
In the KF-DFPC algorithm, the nodes iteratively 
share their KF-updated frequency 
and phase estimates as well as the error covariances with their neighbors, and compute a weighted average of the 
shared values in each iteration to reach a consensus in a CEEC scheme. An existing distributed 
consensus scheme close to CEEC is the CE scheme as used in the diffusion 
KF algorithm (DKF) of \cite{Xin_2022}, 
in which the nodes only share their KF-updated estimates of the electrical 
states with their neighbors, and do not share their error covariances. 
The CE scheme 
reduces the amount of data shared between 
the nodes, but it results in a poorer convergence speed of the DKF algorithm as compared to the CEEC-based KF-DFPC algorithm, 
as shown later in Section \ref{sim_res} of this paper. 
Note that a rapidly 
converging consensus algorithm reduces the latency in achieving the synchronized state of the array and implies fewer exchanges of the data packets between the nodes, 
which reduces energy consumption.

Two other categories of distributed Kalman 
filtering algorithms use either a CM scheme or a CI scheme. In the CM-based KF algorithms proposed in \cite{Saber_2007, Kamgarpour_2008, Sayed_2010}, 
the nodes perform consensus averaging on their locally 
shared measurements and the innovation noise covariance matrices 
to approximate the centralized Kalman filtering algorithm. 
In contrast, in the CI-based KF algorithm of 
\cite{BATTISTELLI_2014}, 
the nodes perform consensus averaging on the KF-predicted information matrix (the inverse of the predicted 
error covariance matrix) and the KF-predicted information vector 
(the information matrix multiplied with the predicted state estimate). 
The CI approach is motivated in \cite{BATTISTELLI_2014} by showing it as the 
solution to minimizing the Kullback-Leibler divergence between the locally 
shared probability density functions (pdfs) and their averaged pdf. 
Thus, combining the interesting features 
of both CM and CI techniques, a hybrid consensus on measurement and consensus on information 
based KF algorithm was proposed in \cite{HCMCI_2015}, which is referred to herein as the KF-HCMCI algorithm, where it was
shown that the KF-HCMCI algorithm 
performs better than distributed KF algorithms that are based only on either the CM or CI scheme. 

Since the CM, CI, and CEEC schemes are complementary, their combination should lead to improved performance in a distributed algorithm. Our approach is based on combining the three schemes into a single distributed Kalman filtering algorithm that obtains better performance at the expense of a slight increase in the required information exchange between nodes in the array.

\subsection{HA-DKF Algorithm}
Kalman filtering is a popular choice for computing 
the minimum mean squared error (MMSE) estimates of the unknown 
parameters if the state transitioning model follows the first-order 
Markov process and the process 
and measurement noises are normally distributed. 
Since these conditions are valid for the frequency and phase evolution models in 
\eqref{FreqPhase_model} and \eqref{FreqPhase_est}, we can use KF at each node 
to obtain the MMSE estimates of these parameters from the local measurements 
\cite{rashid2022frequency}. 
To write the KF equations, we start by defining the state-space model as follows. 
Let the state vector 
$\bdx^n_k=[f_n(k),\theta_n(k)]^T$ represents the unknown state of node $n$ at the 
$k$-th time instant. Then, using \eqref{FreqPhase_model}, we can write the state 
transitioning equation for node $n$ as
\begin{equation}\label{state_eqn}
\bdx^n_k=\bdx^n_{k-1}+\bdu^n_k,
\end{equation}
in which 
$\bdu^n_k\triangleq\left[\delta f_n,\delta \theta^f_n+\delta \theta_n\right]^T$ and 
is modeled as \mbox{$\bdu^n_k\sim\N(\bdzero,\bdQ^n_k)$.} 
The correlation matrix $\bdQ^n_k$ is given by 
\begin{equation}
\bdQ^n_k=\Exp\left[\bdu^n_k\left(\bdu^n_k\right)^T\right]=
\left[
\begin{matrix}
\sigma^2_f && -\pi T\sigma^2_f\\
-\pi T\sigma^2_f && \pi^2T^2\sigma^2_f+\sigma^2_\theta
\end{matrix}
\right].
\end{equation}
Furthermore, we assume that $\bdy^n_k=\left[\hat{f}_n(k),\hat{\theta}_n(k)\right]^T$ 
defines the observation vector of node $n$ 
that combines its frequency and phase estimates, so its observation equation 
can be written as
\begin{equation}\label{obs_eqn}
\bdy^n_k=\bdx^n_k+\bdv^n_k,
\end{equation}
where the vector $\bdv^n_k\triangleq\left[\varepsilon_f,\varepsilon_\theta\right]^T$ 
collects the frequency and phase estimation errors, i.e., 
$\varepsilon_f$ and $\varepsilon_\theta$, respectively. 
Since these errors are normally distributed in \eqref{FreqPhase_est}, 
we can model the measurement noise vector as $\bdv_n\sim\N(\bdzero,\bdSigma_k)$ with the correlation matrix 
$\bdSigma_k$ given by 
\begin{equation}
\bdSigma_k=\Exp\left[\bdv^n_k\left(\bdv^n_k\right)^T\right]=\left[
\begin{matrix}
\left(\sigma^m_f\right)^2 && 0\\
0 && \left(\sigma^m_\theta\right)^2
\end{matrix}
\right].
\end{equation} 

We use the information form of Kalman filtering 
\cite{BATTISTELLI_2014, HCMCI_2015} wherein 
the predicted and the updated 
MMSE estimates of the state at time instant $k$, i.e., $\bdm^n_{k|k-1}$ and 
$\bdm^n_{k| k}$, respectively, 
and their error covariance matrices $\bdV^n_{k|k-1}$ and $\bdV^n_{k|k}$ are translated into the information matrices $\bdOmega^n_{k|k-1}=\left(\bdV^n_{k|k-1}\right)^{-1}$ and $\bdOmega^n_{k|k}=\left(\bdV^n_{k|k}\right)^{-1}$ 
and the information vectors $\bdmu^n_{k|k-1}=\bdOmega^n_{k|k-1}\bdm^n_{k|k-1}$ and 
$\bdmu^n_{k|k}=\bdOmega^n_{k|k}\bdm^n_{k|k}$. Similarly, the noise information 
matrices are defined as $\bdW^n_k=\left(\bdQ^n_k\right)^{-1}$ and 
$\bdU^n_k=\left(\bdSigma_k\right)^{-1}$. 

Now, the HA-DKF algorithm is described as follows. 
To begin, assume that at each time instant $k$, the node $n$ receives the local 
information $\left(\bdmu^m_{k|k-1},\bdOmega^m_{k|k-1},\bdmu^m_{k|k},
\bdOmega^m_{k|k}, \bdy^m_k,\bdU^m_k\right)$ from all $m\in \N_n$ nodes 
where $\N_n$ is the set of neighboring 
nodes of node $n$ including itself\footnote{Since these messages only include the frequency and phase related information, 
they can all be encoded in a small number of data packets 
transmitted between the 
neighboring nodes.}. 
Following~\cite{HCMCI_2015}, in the CM-step of the \mbox{HA-DKF} 
algorithm, the node $n$ computes the weighted average of the scaled 
measurements and the measurement information matrices by 
\begin{align}\label{CM_upd}
\Delta \bdmu^n_k &= \sum_{m\in \N_n} w_{nm}\bdU^m_k\bdy^m_k\nonumber\\
\Delta \bdOmega^n_k &= \sum_{m\in \N_n} w_{nm} \bdU^m_k,
\end{align}
where the weights satisfy $w_{nm}\geq 0$ and \mbox{$\sum_{m\in \N_n}w_{nm}=1$.} 
Further details on defining $w_{nm}$ are discussed later in \mbox{Section \ref{sim_res}.} 
Next in the CI-step, the node $n$ computes the weighted average of the 
locally shared predicted information vectors and information matrices as follows:
\begin{align}\label{CI_upd}
\bdmu^n_k &= \sum_{m\in \N_n} w_{nm} \bdmu^m_{k|k-1}\nonumber\\
\bdOmega^n_k &= \sum_{m\in \N_n} w_{nm} \bdOmega^m_{k|k-1},
\end{align}
At this point, the node $n$ adds up \eqref{CM_upd} and \eqref{CI_upd} to get 
the updated information vector and information 
matrix, i.e., 
\begin{align}\label{CM_CI_correction}
\bdmu^n_{k|k} &= \bdmu^n_k + \Delta \bdmu^n_k\nonumber\\
\bdOmega^n_{k|k} &= \bdOmega^n_k + \Delta \bdOmega^n_k,
\end{align}
\begin{figure*}[t!]
    \begin{minipage}{0.25\textwidth}
     	\includegraphics[width=0.99\textwidth,height=0.65\textwidth]{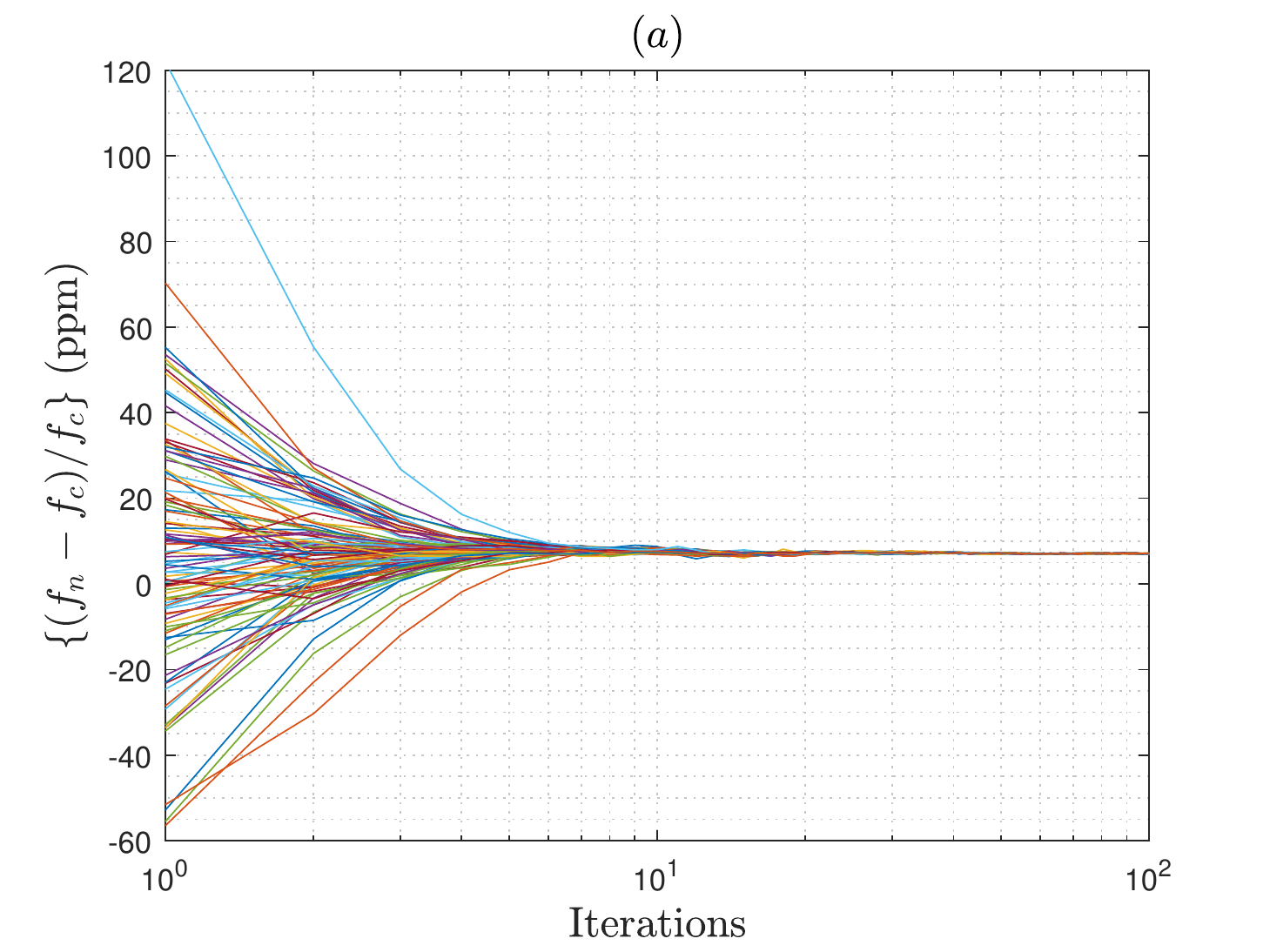}
    \end{minipage}\hspace{.01\linewidth}
    \begin{minipage}{0.25\textwidth}
\includegraphics[width=0.99\textwidth,height=0.65\textwidth]{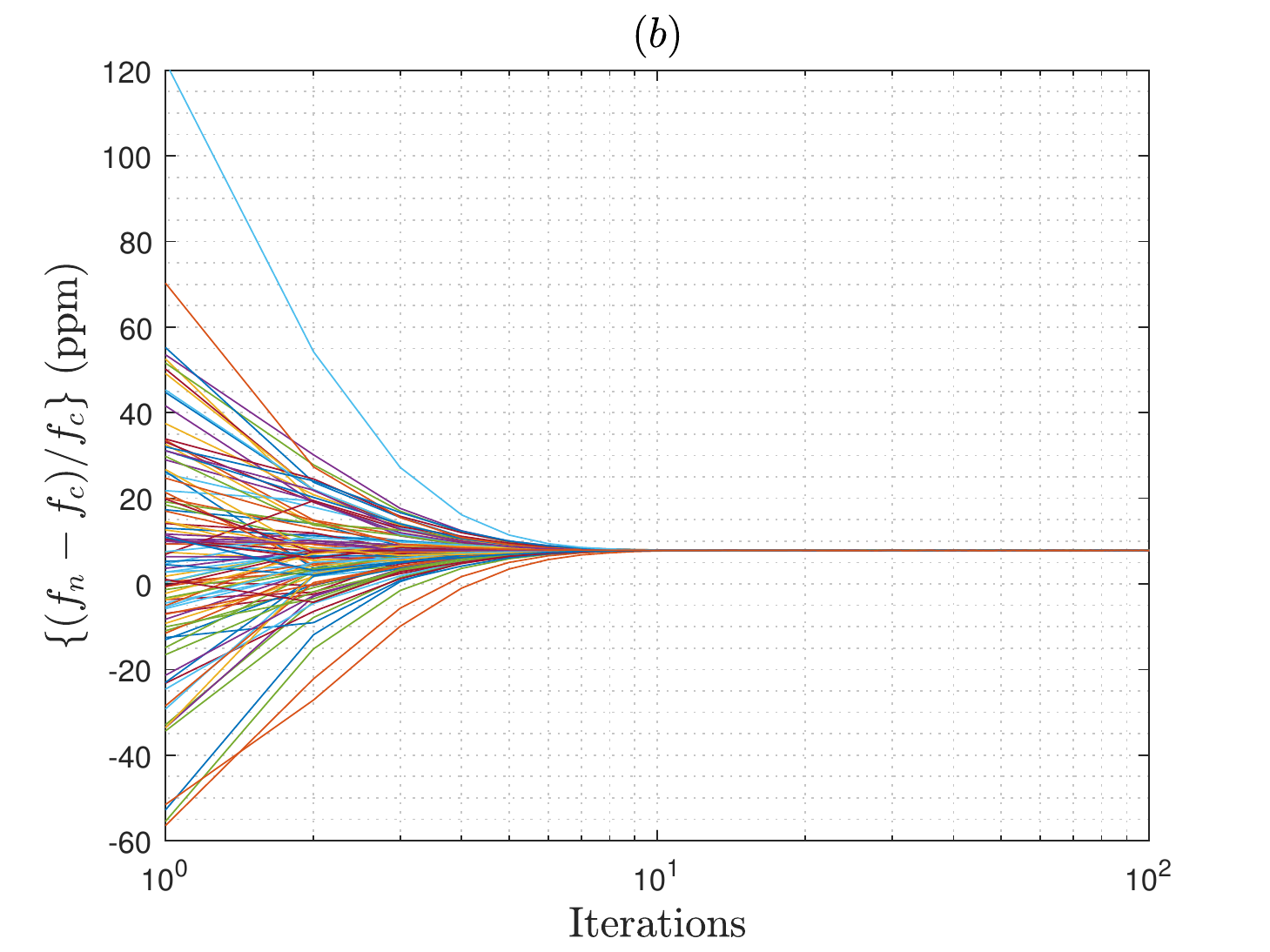}
		\end{minipage}
    \begin{minipage}{0.25\textwidth}
\includegraphics[width=0.99\textwidth,height=0.65\textwidth]{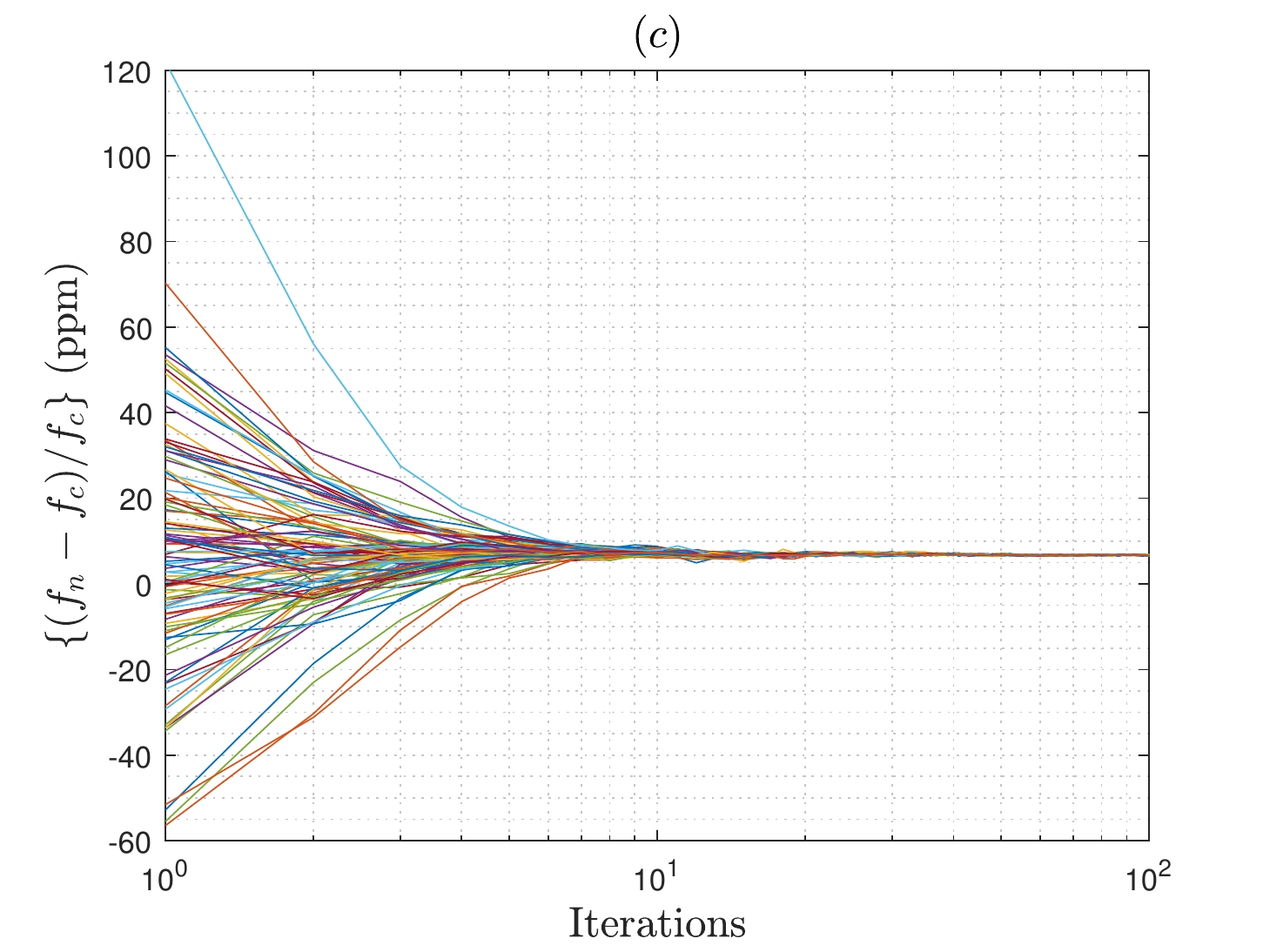}
		\end{minipage}
    \begin{minipage}{0.25\textwidth}
\includegraphics[width=0.99\textwidth,height=0.65\textwidth]{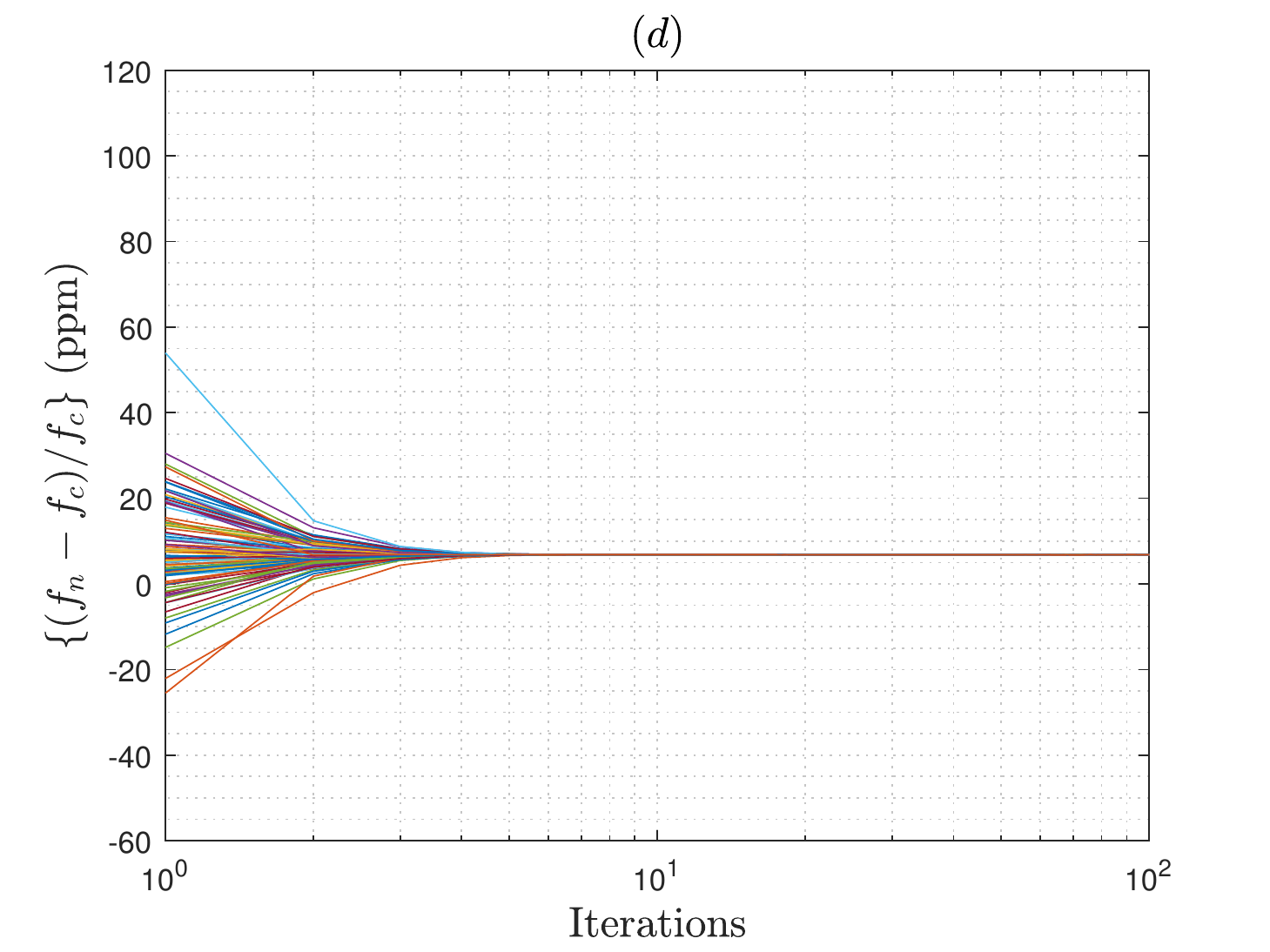}
		\end{minipage}
		\vspace{0.1in}
		\\
    \begin{minipage}{0.25\textwidth}
     	\includegraphics[width=0.99\textwidth,height=0.65\textwidth]{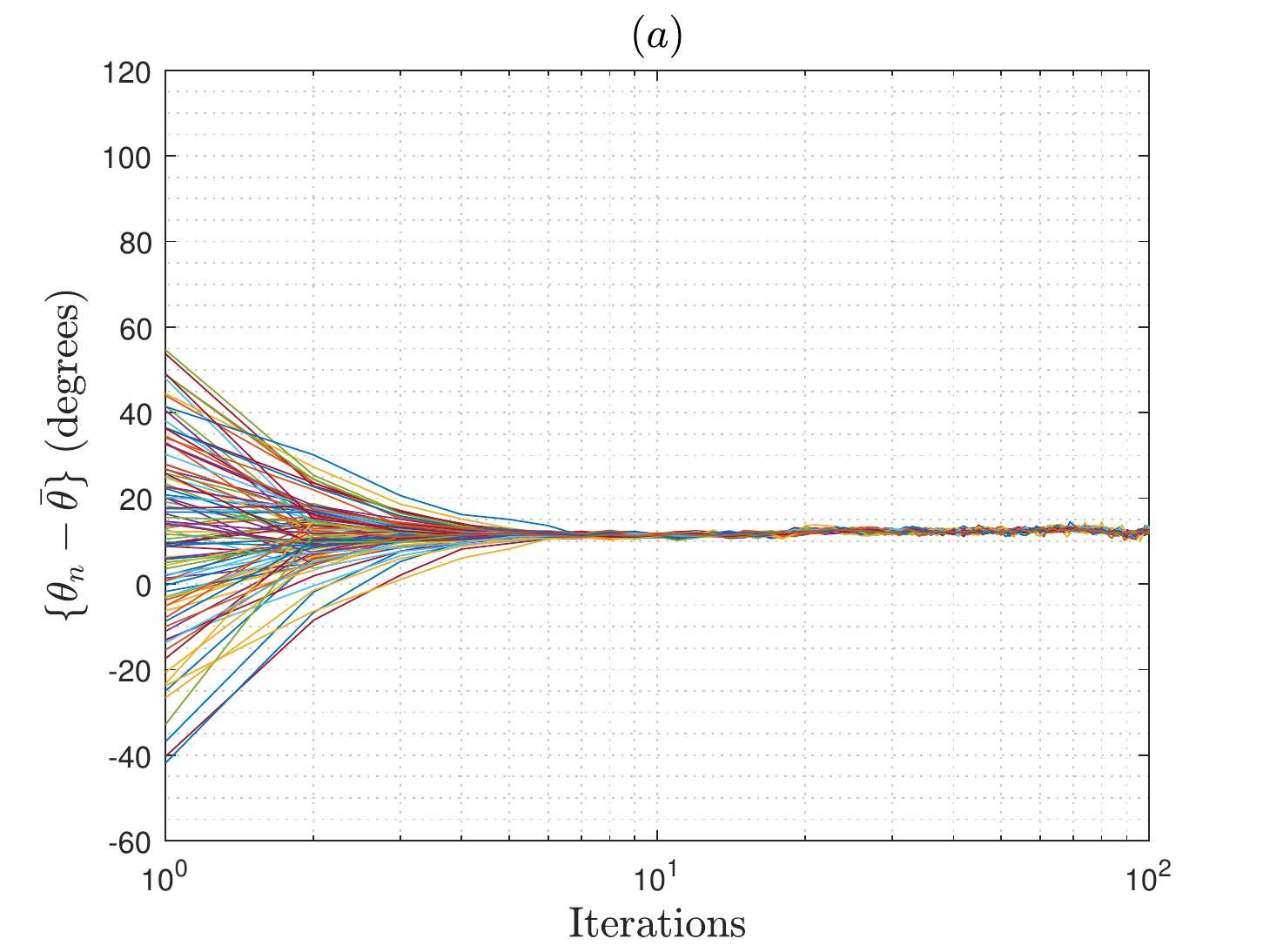}
    \end{minipage}\hspace{.01\linewidth}
    \begin{minipage}{0.25\textwidth}
\includegraphics[width=0.99\textwidth,height=0.65\textwidth]{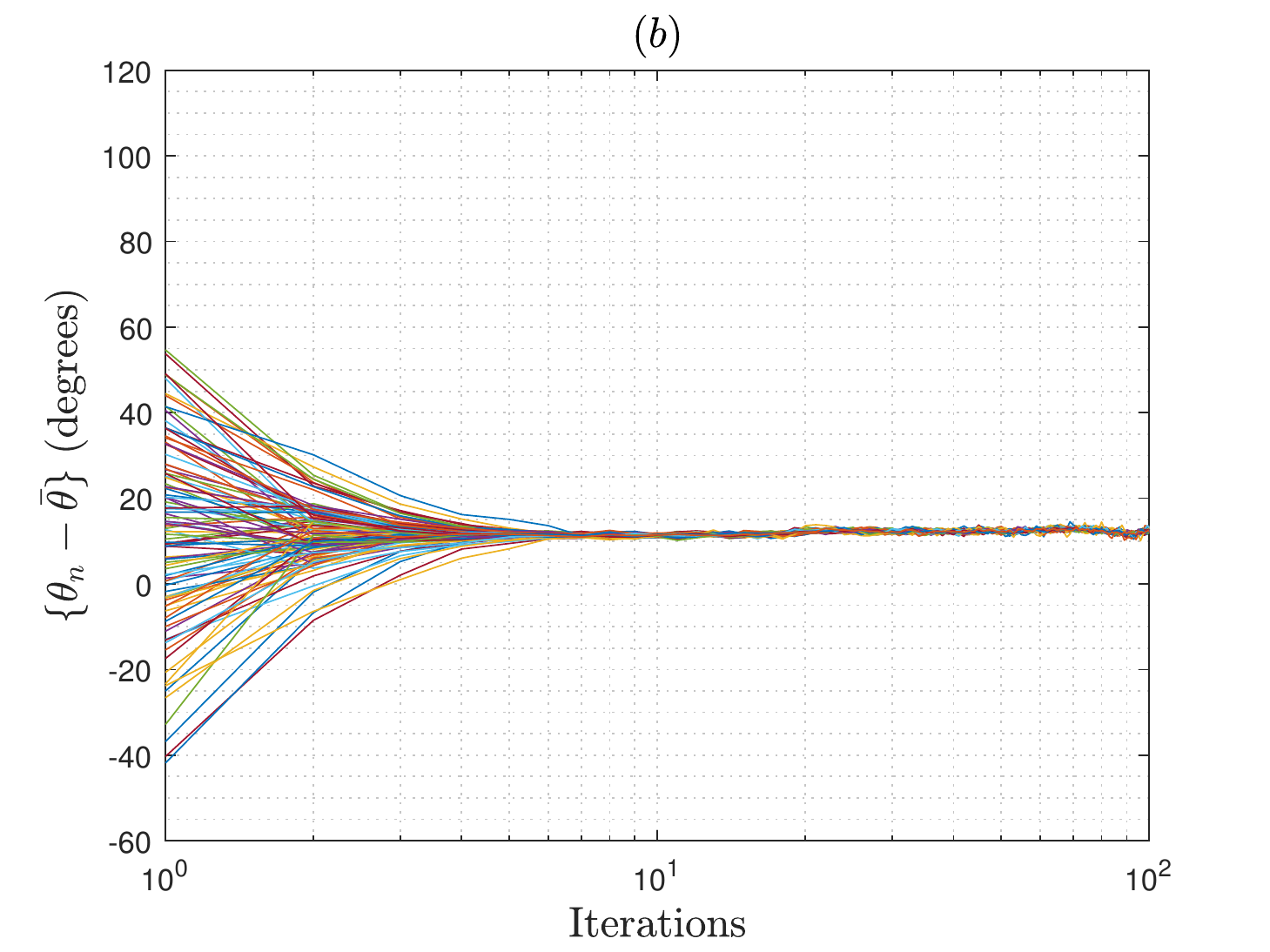}
		\end{minipage}
    \begin{minipage}{0.25\textwidth}
\includegraphics[width=0.99\textwidth,height=0.65\textwidth]{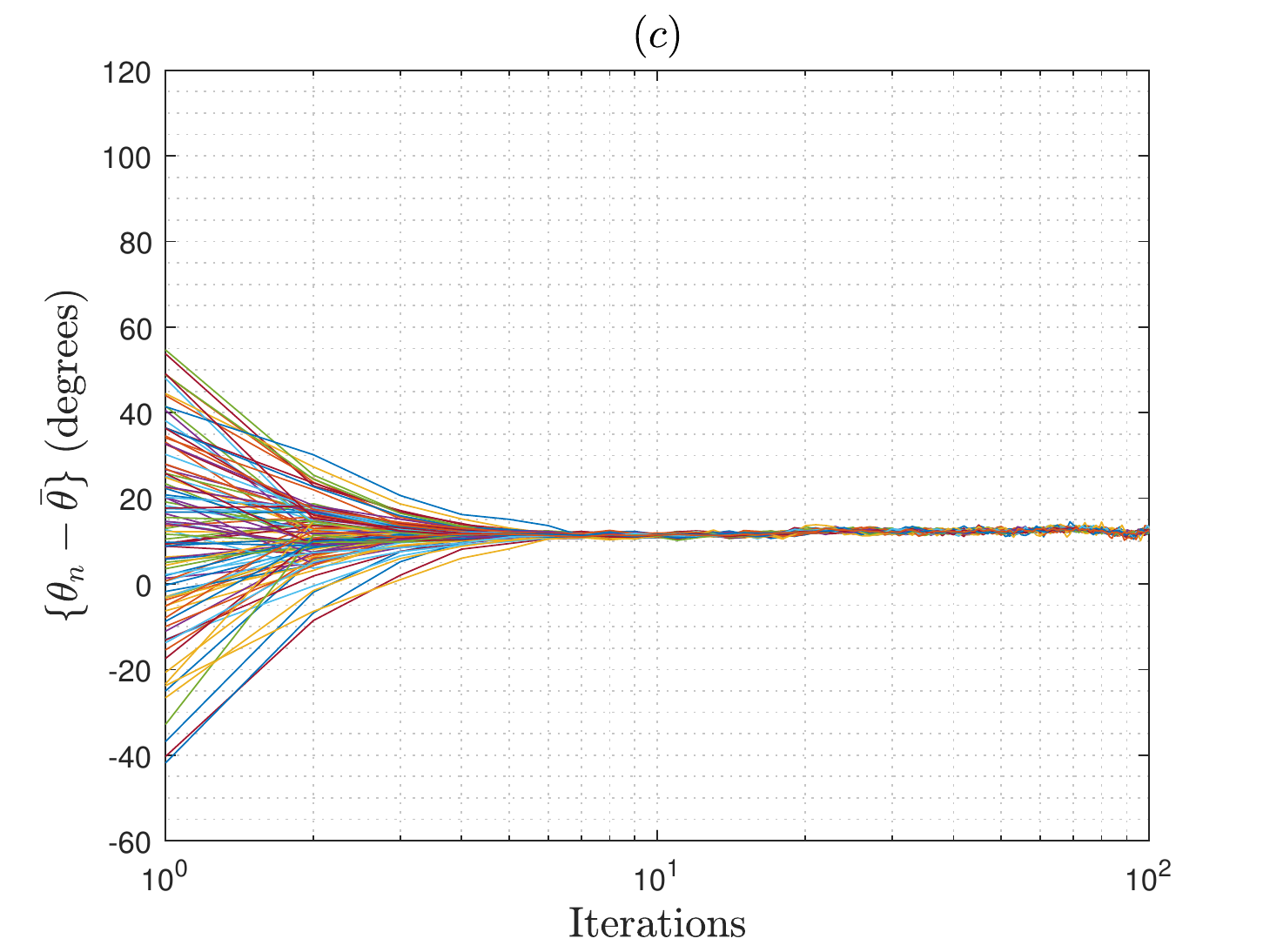}
		\end{minipage}
    \begin{minipage}{0.25\textwidth}
\includegraphics[width=0.99\textwidth,height=0.65\textwidth]{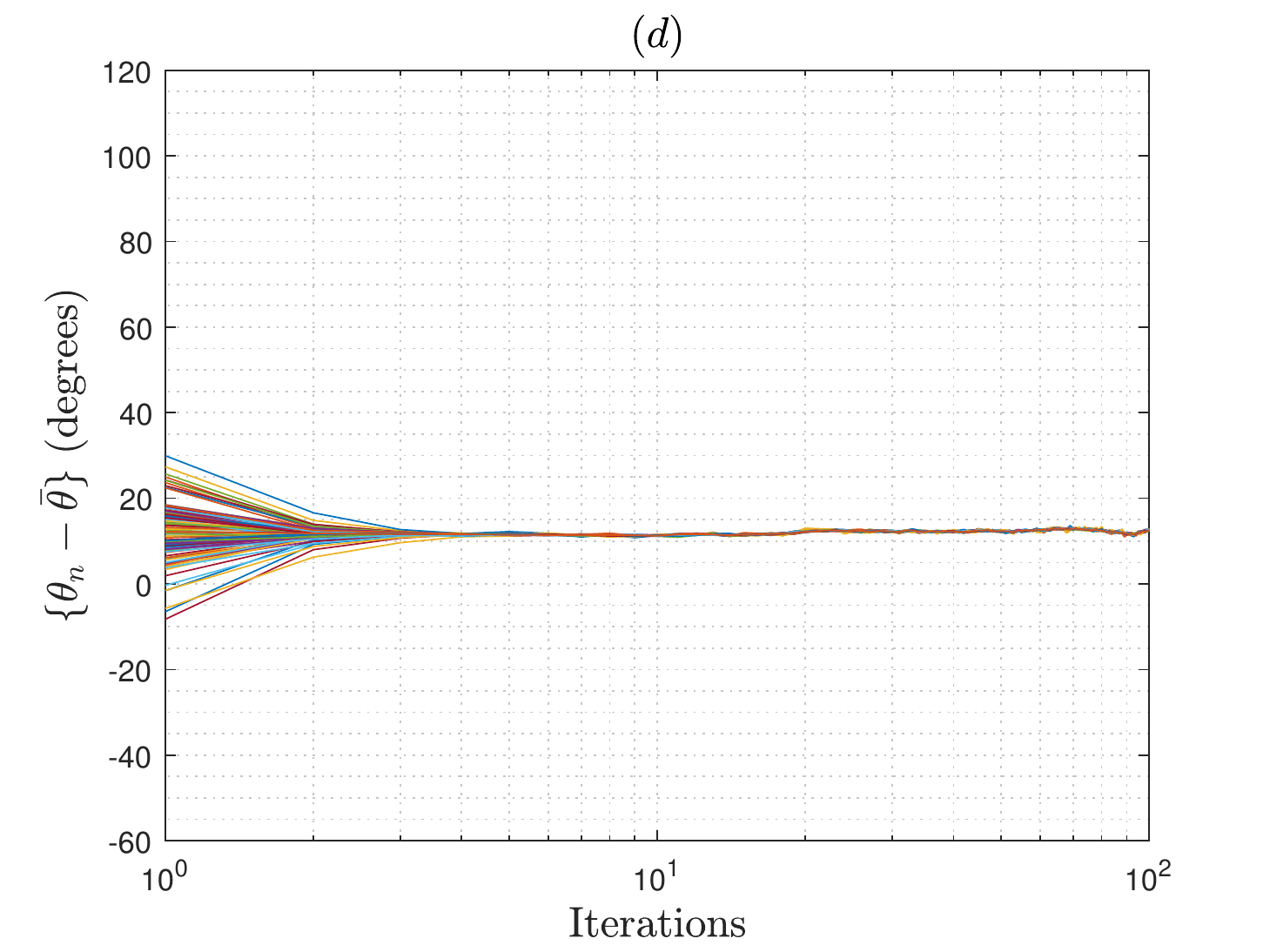}
		\end{minipage}
		\caption{Frequency and phase errors of all the $N$ nodes in the network vs. iterations for $(a)$ DKF, $(b)$ 
		KF-DFPC, $(c)$ KF-HCMCI, and $(d)$ HA-DKF when $c=0.2$, $N=100$, and $SNR=0$ dB}
		\label{fig:freqphase_vs_iter}
\end{figure*}
\begin{figure*}[t!]
		
    \begin{minipage}{0.48\textwidth}
        \centering
\includegraphics[width=0.85\textwidth]{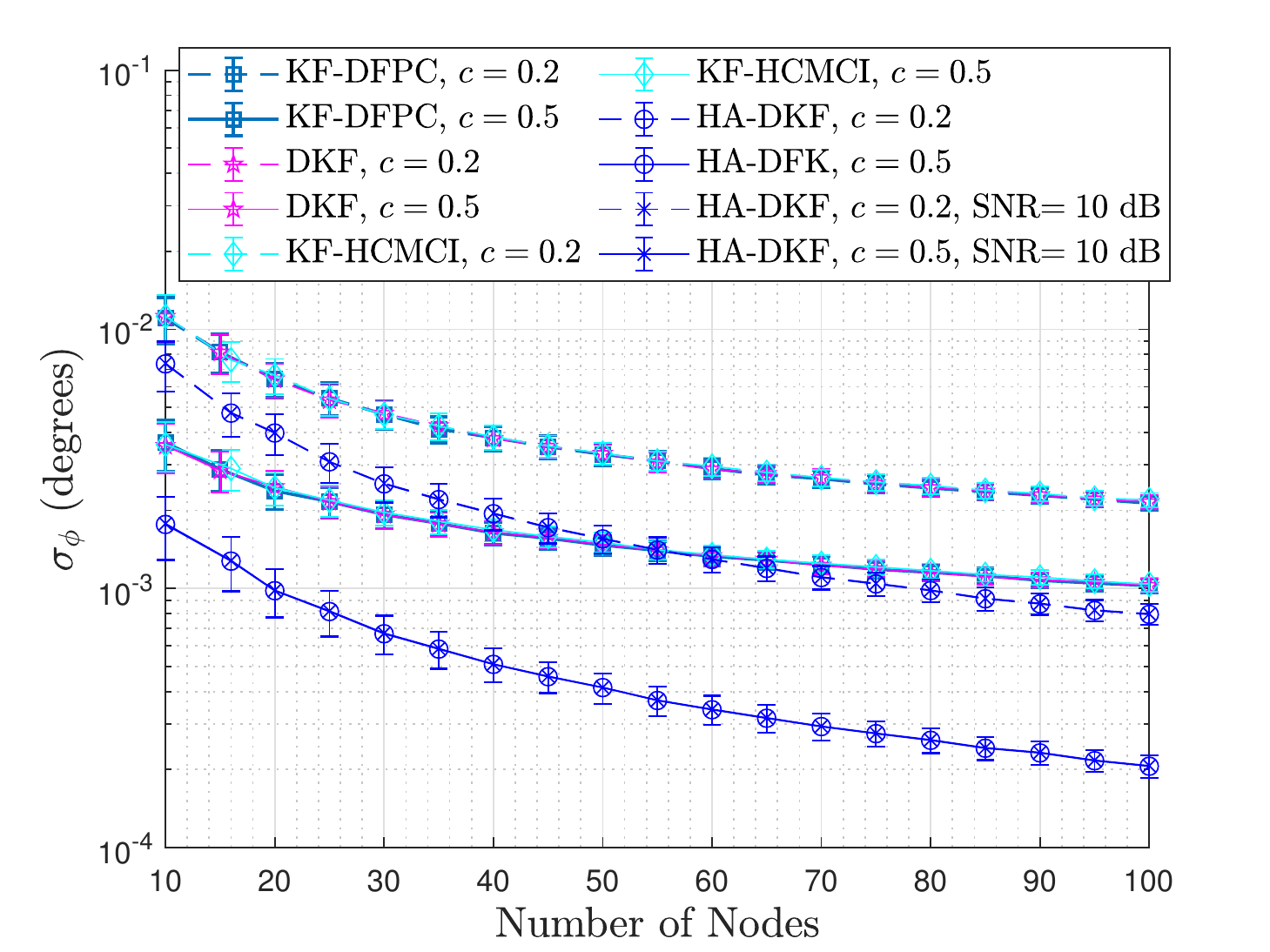}
	\caption{Standard deviation of the total phase error of different algorithms vs. 
	the number of nodes $N$ when SNR$=0$ dB and $c=0.2$ or $0.5$ are used.}
	\label{fig:sigma_vs_N}
		\end{minipage}\hspace{.025\linewidth}
    \begin{minipage}{0.48\textwidth}
        \centering
     	\includegraphics[width=0.85\textwidth]{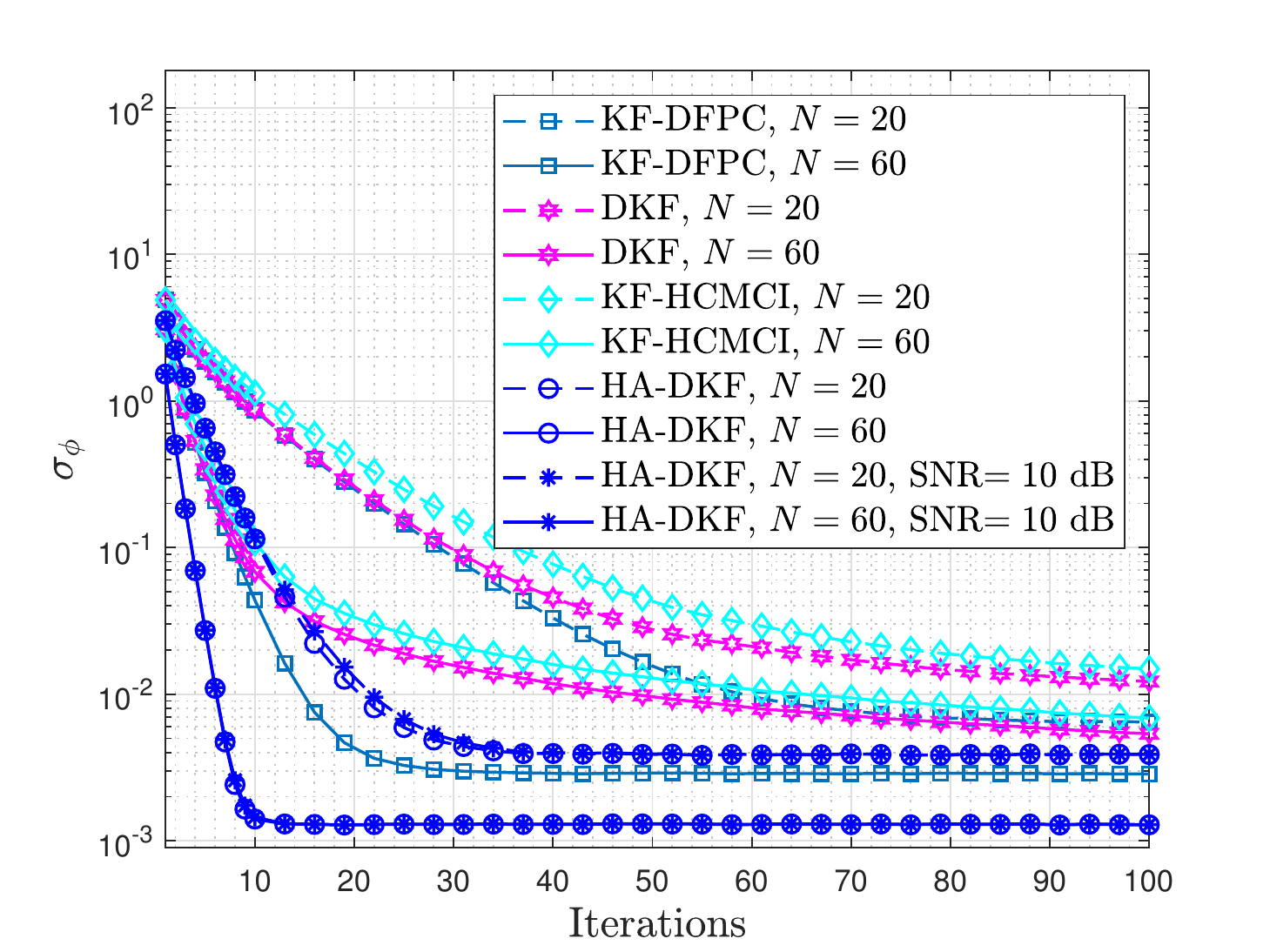}
	\caption{Standard deviation of the total phase error of different algorithms vs. 
	iterations when SNR$=0$ dB, $c=0.2$, and nodes $N=20$ or $60$.}
		\label{fig:sigma_vs_iter}
					\vspace{.2cm}
    \end{minipage}
\end{figure*}

In the CEEC-step of HA-DKF, the node $n$ collects the updated information 
vectors and information matrices from its neighboring nodes $\N_n$ and 
computes the updated state estimate and its error covariance matrix by,
\begin{align}\label{CEEC_upd}
\bdm^n_{k|k} &= \sum_{m\in \N_n} \left(\bdOmega^m_{k|k}\right)^{-1}\bdmu^m_{k|k}\nonumber\\
\bdOmega^n_{k|k} &= 
\left[\begin{matrix}
\sum_{m\in \N_n} |w_{nm}|^2 \nu^{m,f}_{k|k} & 
\sum_{m\in \N_n} |w_{nm}|^2 \nu^{m,f\theta}_{k|k}\\
\sum_{m\in \N_n} |w_{nm}|^2 \nu^{m,f\theta}_{k|k} & 
\sum_{m\in \N_n} |w_{nm}|^2 \nu^{m,\theta}_{k|k}
\end{matrix}\right]^{-1},
\end{align}
where $\nu^{m,f}_{k|k}$, $\nu^{m,f\theta}_{k|k}$, and $\nu^{m,\theta}_{k|k}$ 
in the above equation are the $(1,1)$-th, $(1,2)$-th, and $(2,2)$-th elements of 
the covariance matrix $\left(\bdOmega^m_{k|k}\right)^{-1}$, respectively, 
and the matrix $\bdOmega^m_{k|k}$ is 
obtained from \eqref{CM_CI_correction} for all $m\in \N_n$ 
\cite{rashid2022frequency}. We note that, 
as discussed earlier, the consensus step in \eqref{CEEC_upd} provides a better 
prior distribution 
for the predicted update step of HA-DKF. Thus, using \eqref{CEEC_upd}, 
the predicted information matrix and information vector are computed at 
node $n$ as in \cite{HCMCI_2015} by
\begin{align}\label{pred_eqn}
\bdOmega^n_{k+1|k} &= \bdW^n_{k}-\bdW^n_{k}
\left(\bdOmega^n_{k|k}+\bdW^n_{k}\right)^{-1}\bdW^n_{k}\nonumber\\
\bdmu^n_{k+1|k} &= \bdOmega^n_{k+1|k} \bdm^n_{k|k},
\end{align}
This completes the derivation of the HA-DKF algorithm. 

The computational complexity 
of HA-DKF for node $n$ is dominated by the weighted sum in \eqref{CM_upd} and \eqref{CI_upd}, and by a $2 \times 2$ matrix inversion in \eqref{CEEC_upd} and
\eqref{pred_eqn}. Thus, its complexity is 
$O(|\N_n|+8)$ where $|\N_n|$ is the cardinality of the neighboring nodes 
set $\N_n$ and $O(8)$ are the computations required in inverting a $2\times 2$ matrix. This complexity is the same as that of the KF-DFPC algorithm  
\cite{rashid2022frequency}, the DKF algorithm \cite{Xin_2022}, and the KF-HCMCI 
algorithm \cite{HCMCI_2015}.

\section{Simulation Results}\label{sim_res}
We generated different array networks to 
analyze the synchronization and convergence 
performances of the proposed HA-DKF algorithm. 
For comparison purposes, we use the DKF algorithm \cite{Xin_2022}, the 
KF-DFPC algorithm \cite{rashid2022frequency}, and the KF-HCMCI algorithm with 
$L=1$ consensus step \cite{HCMCI_2015}. All these algorithms are initialized following the initialization of KF-DFPC 
as described in \cite{rashid2022frequency}. The network is randomly generated in each Monte Carlo 
trial with a connectivity $c\in[0.05, 1]$ and 
the weights are defined using the Metropolis-Hastings matrix \cite{rashid2022frequency}. 
The carrier frequency of the nodes is $f_c=1$ GHz, the sampling frequency is $f_s=10$ MHz, and the 
update interval is $T=0.1$ ms. The results were averaged over $10^3$ trials, unless 
stated otherwise.

Fig. \ref{fig:freqphase_vs_iter} shows the frequency and phase errors of DKF, KF-DFPC, KF-HCMCI, and 
the HA-DKF algorithm vs. iterations in a single trial for all the $N$ nodes in the 
network when a moderate connectivity of $c=0.2$ is chosen for $N=100$ nodes in the array and when SNR$=0$ dB 
is used. It is observed that, for all the algorithms, as the iterations increases both frequencies and phases of the nodes 
converge to the average of their initial values; however, the convergence is much faster for the proposed HA-DKF algorithm 
as compared to the other algorithms. 

In Fig. \ref{fig:sigma_vs_N}, we examine the synchronization performances of these 
algorithms vs. varying the number of nodes $N$ in the array. For that purpose, 
we define the total residual phase error of node $n$ as $\delta \phi_n=2\pi \delta f_n T+ 2\pi\varepsilon_f T
+\delta \theta^f_n+\delta\theta_n+\varepsilon_\theta$ and we plot the standard deviation of this phase 
error of all the nodes. 
The connectivity $c$ is either chosen to be $0.2$ 
or $0.5$, and the SNR $=0$ dB is assumed. As observed, the total phase error of all the 
algorithms decreases with an increase in the $N$ and $c$ values. 
This is because the average number of connections per node increases by either increasing $N$ or $c$, 
resulting in a more accurate and stable local average per node. Specifically, our proposed HA-DKF algorithm 
outperforms others in reducing the phase error of the array.
This figure also shows the phase error of HA-DKF when \mbox{SNR $=10$ dB} is assumed. 
As expected, the phase error of HA-DKF is consistent at both lower and higher SNR values.  

Finally, in Fig. \ref{fig:sigma_vs_iter}, we evaluate the convergence performances of these algorithms by 
examining the standard deviation of the total phase error vs. iterations when SNR $=0$ dB, $c=0.2$, and 
the number of nodes are set to either $N=20$ or $60$. Due to the above-mentioned reasons, 
for both $N$ values, our HA-DKF algorithm 
converges faster than the DKF, KF-DFPC, and KF-HCMCI algorithms. As described in Section \ref{Intro-section}, 
the illustrated improvement in the convergence and synchronization performances of our HA-DKF algorithm is 
due to the fusion of all CM, CI, and CEEC schemes in distributed Kalman filtering.
 
\section{Conclusion}\label{conclu}

We investigated the problem of joint frequency and phase synchronization of the nodes 
in a distributed phased array, wherein the oscillators' induced frequency and phase 
offset errors are considered and modeled using practical statistics. A 
consensus averaging based distributed synchronization is considered and an HA-DKF algorithm is proposed in which the nodes 
share their local measurements, the innovation noise covariances, and the KF-predicted and the 
KF-updated estimates and error covariances to significantly reduce the residual phase error of the array in fewer 
iterations. Simulation results show that the HA-DKF algorithm outperforms the DKF, KF-DFPC, and the closely related KF-HCMCI 
algorithm in achieving improved synchronization and convergence performances of the array. 
 
\bibliographystyle{IEEEtran}
\bibliography{References}
\end{document}